\documentstyle[editedvolume,epsfig]{crckapb} 
\def\simlt{\lower.5ex\hbox{$\; \buildrel < \over \sim \;$}}
\def\simgt{\lower.5ex\hbox{$\; \buildrel > \over \sim \;$}}

\begin{opening}

\title{Formation of Bulges} 
\author{Joseph Silk}\author{Rychard Bouwens}
\institute{ Departments of Astronomy and Physics, and Center for Particle
Astrophysics, University of California, Berkeley, CA 94720}
\end{opening}
\runningtitle{BULGES}
\begin{document}
\begin{abstract}
Bulges, often identified with the spheroidal component of a galaxy,
have a complex pedigree.  Massive bulges are generally red and old,
but lower mass bulges have broader dispersions in color that may be
correlated with disk colors. This suggests different formation
scenarios.  I will review possible formation sequences for bulges,
describe the various signatures that distinguish these scenarios, and
discuss implications for the high redshift universe.
\end{abstract}

\section{Three Scenarios for Bulge Formation}

We consider the following possibilities for bulge formation: bulges
form before, contemporaneously with, or after disks.

\underbar{Bulges are old}: 
Monolithic collapse (Eggen, Lynden-Bell and Sandage 1962) described
the formation of population II prior to disk formation.  Evidence on
the age of the inner Milky Way bulge stars generally supports an old
population that formed before the disk.  Galaxies with massive bulges
would have necessarily formed by primordial collapse, major mergers at
high redshifts, or infall of satellite galaxies (Pfenniger 1992).

The chemical evidence is less clear: the predictions of a simple
monolithic formation model cannot be easily reconciled with the
observed abundance spread in bulge stars.  Indeed theory has largely
supplanted a monolithic collapse picture with a clumpy collapse model.
The theory of galaxy formation has had considerable success in
predicting various properties of large-scale structure.  Hierarchical
formation is closer in spirit to the Searle-Zinn view of halo and
bulge formation in which many globular cluster mass clumps merge
together.  We regard bulge and luminous halo formation as closely
related phenomena, the bulge simply being the core of the field star
halo.  Hierarchical galaxy formation involves a sequence of successive
mergers of larger and larger subgalactic scale clumps.  At any given
stage, gas dissipation and settling produces disks that are destroyed
in subsequent mergers.  Not all disk structure is erased: dwarf
satellites and even globular clusters may be substructure relics. The
disk only forms after the last massive merger via gas infall.  In
environments such as rich clusters disk infall is largely suppressed
and the cluster cores are dominated by spheroidal galaxies.  Bulges
are inevitably older than disks, and formed on a dynamical time scale.
Their formation is characterized by a series of intense formation
episodes or starbursts, produced by each merger.  Most stars that are
now in the bulge formed during the process of bulge accumulation.

\underbar{Bulges are of intermediate age}: 
One can also envisage the following prescription for bulge formation.
Merging of dwarf irregular galaxies with a massive disk galaxy will
result in the dwarf being stripped of gas.  The angular momentum of
the gas guarantees that it will eventually dissipate to provide infall
into the disk.  The stellar component, however, as it interacts with
the disk is partially tidally disrupted, to form the thick disk, but
the dense cores undergo dynamical friction, spiraling into the center
to form the bulge. While this may not be appropriate to the inner
Milky Way bulge, such a picture may be relevant to the outer
spheroid. There is evidence for tidal streams that are continuously
generated by disruption of satellites such as the LMC. The age spread
of the globular star clusters is consistent with a model in which
bulges and disks would be of similar age.

\underbar{Bulges are young}:
Bulges may also form slowly by dynamical instabilities of disks.
Secular evolution of disks has occurred in at least some galaxies,
particularly in late-type galaxies (Kormendy 1992; Courteau 1996).
The secular evolution of a cold disk inevitably results in
gravitational instability.  On galaxy scales this is dominated by the
non-axisymmetric modes that induce formation of a bar.  Tidal torques
are expected by the bar on the disk gas, which consequently suffers
angular momentum transfer and forms a massive central concentration.
This in turn eventually tidally disrupts the bar, as well as undergoes
a central starburst and forms the bulge.  This process can repeat as
infall of gas continues and the disk becomes sufficiently massive to
again be gravitationally unstable.  Bar disruption takes up to 100 bar
dynamical timescales.

\section{Signatures of Bulge Formation}

The observed properties of bulges provide a fossil testament to their
formation.  The various signatures do not lead to any unique
conclusion, however.  Consider first the Milky way as a prototype for
bulge formation, bearing in mind that our local neighborhood is
necessarily limited in scope.  The following remarks are largely
summarized from an excellent review of the Milky Way bulge by Wyse,
Gilmore and Franx (1997).

\underbar{Ages}: This should be the cleanest signature.  The Milky Way
bulge appears to be indistinguishable in age from the inner globular
clusters that form a flattened subsystem and appear to be $\sim 2$ Gyr
younger than the oldest globular cluster systems.  However the colors
of other bulges,especially in late-type disks, show a broad dispersion
which may reflect age differences.

\underbar{Abundances}: The observed abundances of Milky Way bulge
stars show a broad dispersion, suggestive of inhomogeneous enrichment,
and the mean metallicity is lower than that of the disk, but higher
than that of the halo.  This supports the inference of an old bulge
from observed ages in localized regions.  In particular the outer
bulge appears to be more metal-poor than the old disk.  If age traces
metallicity, one infers that the bulge precedes the disk.  However
dynamical processes could delay bulge star formation without inducing
chemical evolution.

\underbar{Angular momentum}: The angular momentum distribution of the
inner bulge resembles that of the halo (or outer bulge), rather than
that of the disk.  This is suggestive of a sequential formation
process.

Other bulges provide a broader basis with which to search for clues to
bulge formation.

\underbar{Profiles}: Core radii of bulges and disks are well
correlated.  This suggests that the formation of the two components is
closely coupled.

\underbar{Colors}: Bulge colors show a broad dispersion, but generally
track disk colors.  Both metallicity and age must therefore be
correlated.

\underbar{Dynamics}: Low luminosity bulges are rotationally supported,
as are disks, but luminous bulges generally are supported by
anisotropic velocity dispersion.  Bulges overlap with, but generally
have lower anisotropic velocity dispersion support (i. e.  $\sigma /
v_{circ}$) than do ellipticals.  This suggests that massive bulges are
distinct from disks and closer to ellipticals in dynamical origin,
whereas low luminosity bulges are more closely associated with disk
star formation.

\underbar{Fundamental plane}: Bulges lie in the identical fundamental
plane as ellipticals, although there is a slight offset of the zero
point. This is suggestive of a similar early formation phase for
bulges to that for ellipticals.

The signatures provide mixed signals on the epoch of bulge formation.
It is probably true that many bulges, especially if massive, form
early, while some, especially if associated with later-type spirals,
form late.  The age differences provide an interesting environment
with which to probe bulge formation models.

\section{Bulges at High Redshift}

The high redshift universe potentially provides a unique discriminant.
The differences between the bulge models described above are magnified
at high redshift.

We may broadly classify these bulge formation scenarios into three
types: secular evolution in which bulges form relatively late by a
series of bar-induced starbursts, one in which bulges form
simultaneously with disks, and an early bulge formation model in which
bulges form earlier than disks.  Adjusting the three models to produce
optimal agreement with $z=0$ observations, we compare their
high-redshift predictions with present-day observations, in
particular, with data compiled in various studies based on the CFRS
(Schade et al.\ 1996; Lilly et al.\ 1998) and the HDF (Abraham et al.\
1998).  We have developed a simple scheme for simulating the rival
models and comparing bulge colors and sizes with observations.  Hubble
Space Telescope photometry and color information is available for
galaxy samples that extend up to redshift unity and beyond.  At this
epoch one may hope to detect the difference between old and young
bulge models.

For the purposes of normalizing our models, we examine two local $z=0$
samples (de Jong \& van der Kruit 1994; hereinafter, DJ; Peletier \&
Balcells 1996, hereinafter, PB).  Both are diameter-limited samples
but differ in orientation selection, so that edge-on disks in the PB
sample are simply redder and less prominent relative to the bulges.

Starting with the local properties of disks and a reasonable
distribution of formation times, we construct a fiducial disk
evolution model, to which we add three different models for bulge
formation, the principle difference being simply the time the bulges
form relative to that of their associated disks.  We do not attempt to
model the internal dynamics or structure of spirals (e.g., Friedli \&
Benz 1995).  We adopt the usual Sabc and Sdm luminosity functions
(LFs) for disk galaxies (Binggeli, Sandage, and Tammann 1988).  We
evolve these galaxies backwards in time and in luminosity according to
their individual star formation histories without number evolution.
We take halo formation time to equal the time over which 0.25 of the
final halo mass is assembled.  We normalize by assuming a constant
mass-to-light ratio where $M_{b_J} = -21.1$ corresponds to $4 \times
10^{12} \rm M_{\odot}$ and we adopt the usual CDM matter power
spectrum.  We take star formation in the disk to commence at the halo
formation time with an e-folding time that depends on the $z=0$ galaxy
luminosity, to roughly fit the $z=0$ colour-magnitude relationship.
We assume exponential profiles for the disks with a $b_J$ central
surface brightness modified to account for the observed correlation
between surface brightness and luminosity (e.g., de Jong 1996; McGaugh
\& de Blok 1997).  We compute bulge spectra for the purposes of
determining colours and magnitudes using the Bruzual \& Charlot
instantaneous-burst metallicity-dependent spectral synthesis tables
(Leitherer et al.\ 1996).

To calibrate our fiducial disk evolution models, we compare the model
predictions to both the colour-magnitude relationship of disks in
spirals and the cosmic history of luminosity density.  There is good
agreement with the colour-magnitude relationship. All models, for
which bulge, disk, and E/S0 contributions have been considered,
produce fair agreement with the luminosity density of the universe at
all redshifts for which observable constraints are available (Lilly et
al.\ 1996; Madau et al.\ 1996; Connolly et al.\ 1997).

If bulges form through the merging of disk galaxies, the formation of
the stars found in bulges is expected to precede the formation of
stars in the disks which form out of gas that accretes around the
spheroid ({\it e.g.}, Kauffmann \& White 1993; Frenk {\it et al.}\
1996).  For simplicity, we commence star formation in the bulge 4 Gyr
prior to the formation of disks in our fiducial model and suppose that
it continues for $\tau_{burst} = 0.1$ Gyr.

Suppose next that star formation in the bulge commences at the
formation time of disks, for example because high angular momentum gas
forms the disk while low angular momentum gas simultaneously forms the
bulge which undergoes a mild starburst.  One can imagine gas-rich
satellite infall.  The gas is tidally stripped and accreted onto the
disk at large radii, whereas the dense cores lose angular momentum by
dynamical friction and are incorporated into the bulge.  A refinement
of this model would allow for a sequence of early mergers that formed
the bulge. However the final merger dominates the luminosity and
therefore the spectral evolution, since the star formation efficiency
is greatest for the most massive systems.

Finally, in secular evolution, bulges form after disks, with gas
accretion onto the disk triggering the formation of a bar that drives
gas inflow into the center followed by star formation (Friedli \& Benz
1995).  The build-up of a central mass destroys the bar and inhibits
gas inflow (Norman, Sellwood, \& Hasan 1996), consequently stopping
star formation in the bulge until enough gas accretes onto the galaxy
to trigger the formation of a second bar, followed by a second central
starburst.  Somewhat arbitrarily, we suppose that the first central
starbursts occur some 2 Gyr after disk formation in our fiducial
model, that central starbursts last $\tau_{burst}$ = 0.1 Gyr, as in
the simulations by Friedli \& Benz (1995), and that 2.4 Gyr separate
central starbursts, in order to illustrate the general effect of a
late secular evolution model for the bulge.  We assume that the star
formation rate follows an envelope with an e-folding time equivalent
to the history of disk star formation.  We thereby force star
formation in the disk and the bulge to follow very similar time
scales, given the extent to which they are both driven by gas infall
processes.

We add a simple model for E/S0 galaxies to aid with the interpretation
of observed high redshift, high $B/T$ systems, somewhat arbitrarily
assuming that the distribution of formation redshifts for the $E/S0$
population is scaled to be at exactly twice the distribution of
formation redshifts.

We perform all our calculations using a galaxy evolution software
package written by one of the authors for calculating how the gas,
metallicity, star formation, luminosity, and colours vary as a
function of time for a wide variety of morphological types, formation
times, and star formation histories.  With this software package, we
present representative HDF simulations for our three bulge formation
models in Figure 1 for comparison with the observations.

Clearly the secular evolution model, with late bulge formation, has a
paucity of large B/T objects relative to the other models (Figure 2).
The simultaneous bulge formation model has a large number of such
galaxies simply because a large number of bulges were forming at this
time, while the early bulge formation model has a slightly lower value
due to the fact that bulges in this model had long been in place
within their spiral hosts.

As expected, in all redshift bins, bulges are slightly bluer in the
late bulge formation models than are the disks (Figure 3).  A blue
tail may be marginally detectable in the Schade {\it et al.} data in
the highest redshift bins.  Unfortunately, given the extremely limited
amount of data and uncertainties therein, little can be said about the
comparison of the models in all three redshift bins, except that the
range of bulge and relative bulge-to-disk colours found in the data
appears to be consistent with that found in the models.

While consistent with currently available data, our models for bulge
formation are schematic and are intended to illustrate the observable
predictions that will eventually be made when improved data sets are
available in the near future. Our models are still quite crude,
assuming among other things that the effects of number evolution on
the present population of disks can be ignored to $z \sim 1$ as
suggested, for example, in Lilly {\it et al.} 1998.  In contrast, one
recent analysis (Mao, Mo, \& White 1998) has argued that observations
favor the interpretation that a non-negligible amount of merging has
taken place in the disk population from $z=0$ to $z=1$.  For this
particular interpretation, it remains to be seen how all the present
stellar mass in disks could have built up if disks were continually
destroyed by merging to low $z$ given the constraints on the cosmic
star formation history.  Infall of metal-poor gas provides a
non-destructive alternative that is supported by chemical evolution
modeling of the old disk and even by observations of a reservoir of
high velocity outer halo clouds.

\section*{Acknowledgements}

We thank our collaborator Laura Cayon for many inspiring discussions of bulge issues.
This research has been supported in part by NSF.

\section{References}
\def\mnras{MNRAS}
\def\araa{ARAA}
\def\apj{ApJ}
\def\aj{AJ}
\def\pasp{PASP}
\def\apjl{ApJ}

{}

\newpage

\def\eps@scaling{.95}
\def\plotone#1{\centering \leavevmode
\epsfxsize=\eps@scaling\columnwidth \epsfbox{#1}}

\begin{figure}
\resizebox{15cm}{!}{\includegraphics*[27,127][576,666]{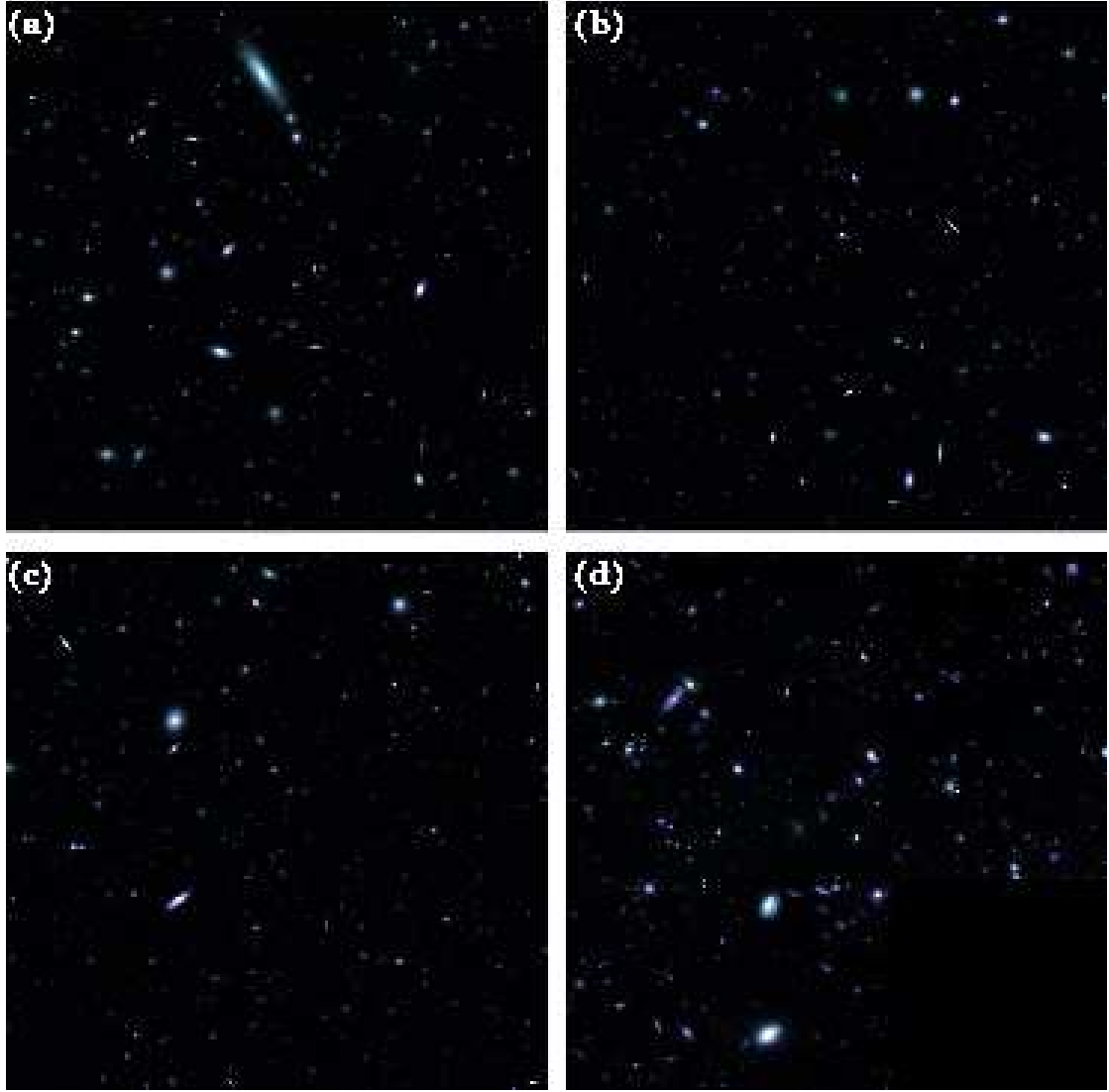}}
\caption{Comparison of simulated BVI images of a 2'' x 2'' patch of
the HDF with the observed images (panel d).  Panel (a) illustrates our
secular evolution model for bulges, panel (b) illustrates our
simultaneous formation model, and panel (c) illustrates our early
bulge formation model.  Calculations are performed using a
galaxy-evolution software package written by one of the authors.}
\end{figure}

\newpage
\begin{figure}
\plotone{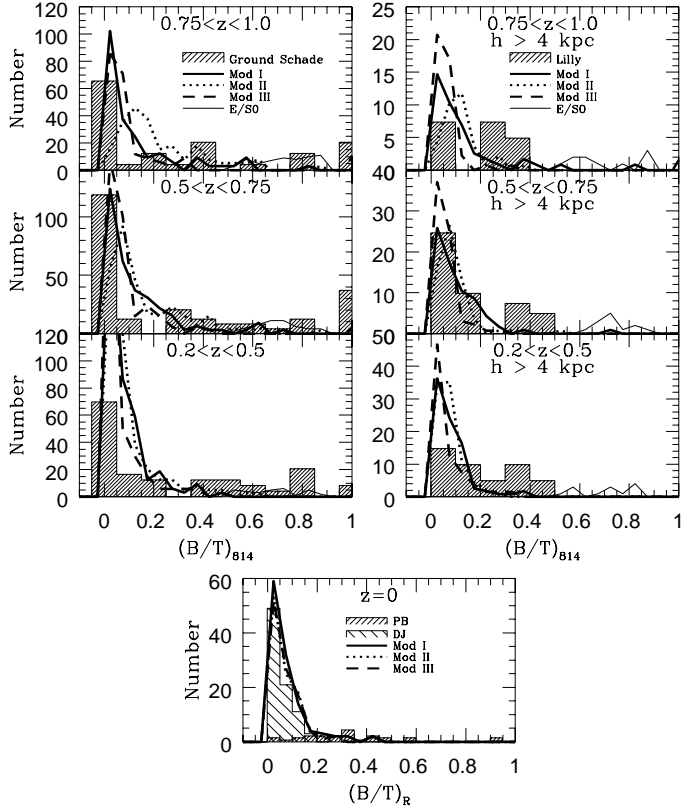}
\caption{Comparison of the observed bulge-to-total ratios (histograms)
with three bulge formation models: a secular evolution model (Mod I,
solid line), a simultaneous formation model (Mod II, dotted line), and
an early bulge formation model (Mod III, short dashed line) (from
Bouwens, Cayon and Silk 1999).  High
redshift comparisons are performed in the upper left panels against
the Schade et al.\ (1996) data using the CFRS selection criteria and
in the upper right panels against the Lilly et al.\ (1996) data using
the CFRS selection criteria plus a size cut ($h > 4 kpc$).  $E/S0$
predictions are also included in the high redshift figures (long
dashed line).  Models are renormalized to match the data.  The de Jong
\& van der Kruit (1994) and Peletier \& Balcells (1996) samples are
used for the low-redshift comparisons.}
\end{figure}

\newpage

\begin{figure}
\plotone{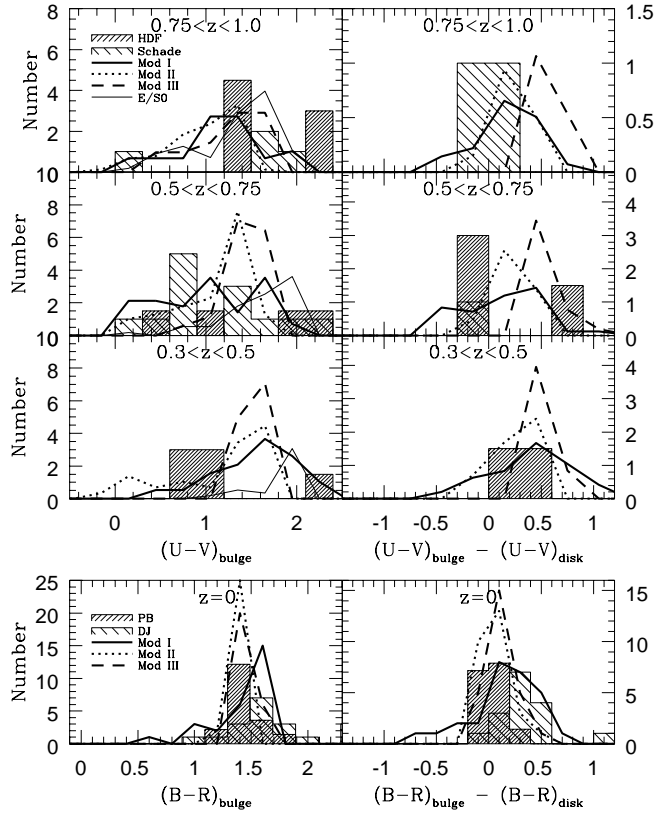}
\caption{Comparison of the observed bulge and relative bulge-to-disk
colours (histograms) with those of the models, at both high and low
redshift (from
Bouwens, Cayon and Silk 1999).  Model curves (renormalized to match observations and
multiplied by 1.6 to increase their prominence) and low redshift data
are represented as in Figure 2.  The high redshift comparison includes
data from the HDF for the Bouwens et al.\ (1998) sample (shaded
histogram) and HST data from Schade et al.\ (1995) (open histogram).}
\end{figure}

\end{document}